\documentclass[12pt,a4paper]{article}

\usepackage{latexsym}
\usepackage{amsmath}
\usepackage{amsfonts}
\usepackage{amssymb}

\newcommand{\be}{\begin{equation}}
\newcommand{\ee}{\end{equation}}
\newcommand{\bea}{\begin{eqnarray}}
\newcommand{\eea}{\end{eqnarray}}

\def\Ad{{\mathrm{Ad}}}                  %
\def\H{{\cal H}}                        %
\def\End{{\mathrm{End}}}                %
\def\cO{{\cal O}}                       %
\def\K{{\cal K}}                        %
\def\G{{\cal G}}                        %
\def\B{{\cal B}}                        %
\def\cA{{\mathcal A}}                   %
\def\ext{\mathrm{ext}}                  %
\def\red{\mathrm{red}}                  %
\def\cI{{\mathcal I}}                   %
\def\Diff{\mathrm{Diff}}                %
\def\Lie{\mathrm{Lie}}                  %
\def\cL{{\cal L}}                       %
\def\M{{\mathcal M}}                    %
\def\Fun{{\mathrm{Fun}}}                %
\def\F{{\cal F}}                        %
\def\eff{{\mathrm{eff}}}                %
\def\bR{{\mathbb R}}                    %
\def\bZ{{\mathbb Z}}                    %
\def\cJ{{\cal J}}                       %
\def\rad{{\mathrm{rad}}}                %
\def\g{{\mathbf{g}}}                    %
\begin{document}
\setcounter{page}{0}
\thispagestyle{empty}

\vspace*{0.5cm}
\begin{center}
{\Large \bf
Hamiltonian reductions of free particles
under polar actions of compact Lie groups }

\end{center}

\vspace{0.2cm}

\begin{center}
L. FEH\'ER${}^{a}$ and B.G. PUSZTAI${}^b$ \\

\bigskip

${}^a$Department of Theoretical Physics, MTA  KFKI RMKI\\
1525 Budapest 114, P.O.B. 49,  Hungary, and\\
Department of Theoretical Physics, University of Szeged\\
Tisza Lajos krt 84-86, H-6720 Szeged, Hungary\\
e-mail: lfeher@rmki.kfki.hu

\bigskip
${}^b$Centre de recherches math\'ematiques, Universit\'e de Montr\'eal\\
C.P. 6128, succ. centre ville, Montr\'eal, Qu\'ebec, Canada H3C 3J7, and\\
Department of Mathematics and Statistics, Concordia University\\
1455 de Maisonneuve Blvd. West, Montr\'eal, Qu\'ebec, Canada H3G 1M8\\
e-mail: pusztai@CRM.UMontreal.CA

\end{center}

\vspace{0.2cm}

\begin{abstract}
Classical and quantum Hamiltonian reductions of free geodesic systems of complete
Riemannian manifolds are investigated.
The reduced systems are described
under the assumption that the underlying compact symmetry group acts in a polar manner
in the sense that there exist regularly embedded, closed, connected  submanifolds meeting all
orbits orthogonally in the configuration space.
Hyperpolar actions on  Lie groups and
on symmetric spaces lead  to families of integrable systems of spin Calogero-Sutherland type.

\end{abstract}

\newpage
\section{Introduction}
\setcounter{equation}{0}

In the theory of integrable systems one of the basic facts is that
many interesting models arise as Hamiltonian reductions of certain
canonical `free' systems that can be integrated `obviously' due to
their large symmetries. For example, the celebrated Sutherland model
of interacting particles on the line, defined classically by the
Hamiltonian
\be
\H(q,p) = \frac{1}{2} \sum_{k=1}^n p_k^2 +
\nu (\nu -1) \sum_{1\leq i<j\leq n} \frac{1}{\sin^2(q_i-q_j)},
\label{1.1}\ee
can be viewed as a reduction of the canonical geodesic system on the
group $SU(n)$. The underlying symmetry is given by the conjugation
action of the group on itself, and the  model (\ref{1.1}) results
(in the center of mass frame) by fixing the Noether charges of this
symmetry in a very special manner. The Hamiltonian reduction can be
performed both at the classical \cite{KKS} and at the quantum
mechanical \cite{EFK} level. For reasons of representation theory,
in the latter case one obtains the model with integer values of the
coupling constant $\nu$. The derivation of the Sutherland model
(\ref{1.1}) by Hamiltonian reduction has been generalized to obtain
other  integrable models of (spin) Calogero-Moser-Surtherland type
\cite{OPI,OPII,GH,Nekr,Res,AKLM,Obl,Ho,FP1,FP2,FP3}.

The most direct generalizations \cite{Nekr,Res} result by
replacing $SU(n)$ with an arbitrary connected, compact simple Lie group,
and fixing the Noether charges arbitrarily.
Then the reductions relying on the conjugation symmetry lead to spin Sutherland type models
with interaction potentials determined by the corresponding root system.
Classically, the phase spaces of these models contain additional (`spin')
degrees of freedom besides the cotangent bundle of the reduced configuration space
representing particle coordinates. Quantum mechanically, the Hamiltonian acts
on multicomponent wave functions as opposed to the scalar wave functions
of the spinless Sutherland type models.
These wave functions correspond to  vector valued spherical functions on the symmetry group,
 much studied in harmonic analysis.

Other generalizations start from geodesics systems on symmetric spaces.
The pioneering contributions in this line
of research are due to Olshanetsky ad Perelomov \cite{OPI,OPII}.
For recent developments, see e.g.~\cite{FP1,FP2,FP3} and references therein.

In the examples alluded to the starting point is always a connected, complete Riemannian manifold
equipped with an isometric action of a compact symmetry group that admits \emph{flat sections}.
A \emph{section}, in the sense of Palais and Terng \cite{PT},
for an isometric action of a group $G$ is a regularly embedded, closed, connected
submanifold that meets
each $G$-orbit at least once and it does so orthogonally at every intersection point with an orbit.
An action that admits sections is called \emph{polar}, and the polar actions with  \emph{flat}
sections are called \emph{hyperpolar}.
For example, the conjugation action of a connected compact Lie group on itself is hyperpolar,
with the sections
being the maximal tori.
Polar actions, especially on symmetric spaces, have been much studied
in the literature \cite{Con,Sze,HPTT,Kol}.

The aim of this paper is to present some general results
concerning Hamiltonian symmetry reduction under polar actions of compact Lie groups.
Our motivation comes from our interest in
integrable systems, for which  examples of such reductions have already proved useful as
mentioned above.
On the one hand, we realized that
both the classical and the quantum Hamiltonian reduction can be easily treated
in a unified manner in the setting of polar actions.
On the other hand, it turns out that the reduced systems arising from
hyperpolar actions on Lie groups and on symmetric spaces
yield, generically,  spin Calogero-Sutherland type integrable models.
This article may serve as a  step towards exploring these models systematically.

The organization of the paper and the main results  are as follows.
After recalling some  background material in Chapter 2, we deal with the
classical Hamiltonian reduction of the geodesic motion focusing
on the dense, open submanifold of regular elements in the configuration space.
The main result in Chapter 3 is Theorem 3.1, which characterizes the reduced system
arising at an arbitrary value of the momentum map. This result may be obtained also
by specializing more general results of Hochgerner \cite{Ho,Ho+} on reduced cotangent bundles.
We give a simple, self-contained proof of it directly in the framework of polar actions.
Chapter 4 is devoted to quantum Hamiltonian reduction. In essence, this amounts to
restricting the Laplace-Beltrami operator of the  Riemannian manifold to certain
equivariant wave functions.  Here our main result is Theorem 4.5, stating both the explicit form
of the reduced Laplace-Beltrami operator and its self-adjointness on a suitable domain.
In Chapter 5 we collect some  examples of hyperpolar actions,
which lead to spin Calogero-Sutherland type models as will be detailed elsewhere.

Although some of the results that we derive, or their special cases,  are already known,
to our knowledge they have not yet been  treated from the unifying point of view of polar actions.
We thought it worthwhile to systematize and extend the existing results  from this perspective.
An important feature of our work is that we present the classical and
quantum Hamiltonian reduction as two sides of the same coin, showing both the similarities
and the differences between the corresponding classical and quantum reduced systems.

\section{Basic assumptions and definitions}
\setcounter{equation}{0}

Let us consider a smooth, connected, complete Riemannian manifold $Y$ with metric $\eta$ and a
compact Lie group $G$ that acts as a symmetry group of $(Y,\eta)$.
It is a very nice situation when the group action admits {\em sections} as defined and
described in detail by Palais and Terng \cite{PT}.
By definition,  a section $\Sigma$ is a connected, closed, regularly embedded,
smooth submanifold of $Y$
that meets every $G$-orbit and it does so orthogonally at every
intersection point of $\Sigma$ with an orbit.
As was already mentioned, the group actions permitting sections are called \emph{polar},
and \emph{hyperpolar}
if the sections
are flat in the induced metric.
(Some authors relax the definition in various ways, e.g.~the section is sometimes
required to be an immersed submanifold only.)

Denote by $\check Y$ the open, dense, $G$-invariant submanifold of $Y$
consisting of the $G$-regular points of principal orbit
type\footnote{We call the points of $\check Y$ simply  `regular'.
For a general reference on group actions, see e.g.~\cite{GOV}.
Note also that in the notion of \emph{manifold} we
include the second axiom of countability.}, i.e., those points
$y\in Y$ whose isotropy subgroups, $G_y$, are the smallest possible for the given $G$-action,
$G\ni g \mapsto \phi_g \in \Diff(Y)$.
Since the isotropy subgroups of all elements of $\check Y$ are conjugate in $G$,
the space of orbits $\check Y/G$ is a smooth manifold, carrying a naturally
induced Riemannian metric that we call $\eta_{\red}$.
For a section $\Sigma$,
denote by $\check \Sigma$ a connected component of the manifold
 $\hat \Sigma := \check Y \cap \Sigma$.
It is known that $\check Y/G$ is connected, and the restriction of the
natural projection $\pi: \check Y \to \check Y/G$ to $\check \Sigma$ is an isometric
diffeomorphism, with the metric on $\check \Sigma$ induced from
its being a submanifold of $Y$.
The isotropy subgroups of all elements of $\hat \Sigma$ are the same, and for a
fixed section we define  $K:= G_y$ for $y\in \hat \Sigma$.

It is also worth noting that if a section exists, then
there is a unique section through every $y\in \check Y$, which is the image of the
orthogonal complement of $T_y(G.y)$ in $T_y Y$ by means of the geodesic exponential map.
Moreover, one has the generalized Weyl group $W(\Sigma)= N(\Sigma)/K$, where
$N(\Sigma)$ is the subgroup of $G$ consisting of the elements that map
$\Sigma$ to $\Sigma$, including $K$ as a normal subgroup of finite index.
The group $W(\Sigma)$ permutes the connected components of $\hat \Sigma$.

A very simple example is the standard action of
$SO(2)$ on the plane for which the sections are the straight lines through the origin.
In this case $\Sigma \setminus \hat \Sigma$ is just the origin,
and $W\simeq {\bZ}_2$. The term `polar' action reminds one of the
availability of `polar coordinates'.
Many interesting examples can be found e.g.~in \cite{HPTT,Kol}.

For our purpose, we choose a $G$-invariant, positive definite scalar product,
$\B$, on $\G:= \Lie(G)$ and often identify $\G$ with its dual  $\G^*$ by means of $\B$.
We have the Lie algebras  $\G_y:= \Lie(G_y)$,
in particular $\K:= \Lie(K)$,
and the corresponding
orthogonal decompositions,
\be
\G= \G_y \oplus \G_y^\perp,
\qquad \forall y\in Y
\label{2.1}\ee
induced by $\B$.
These give rise to the identifications
$(\G_y)^*\simeq \G_y$, $(\G_y^\perp)^* \simeq \G_y^\perp$.
We decompose the tangent space $T_y Y$ into vertical and horizontal subspaces as
\be
T_y Y= V_y \oplus H_y
\quad\hbox{with}\quad
V_y := T_y (G.y),
\quad
H_y := V_y^\perp,
\label{2.2}\ee
where orthogonality is defined by $\eta_y$.
By using $\eta_y$, we identify $T_yY$ with its dual $T_y^* Y$, whereby
$V_y^* \simeq V_y$ and $H_y^*\simeq H_y$.
At the regular points we also have
\be
H_y  =T_y \hat \Sigma = T_y \Sigma
\qquad
\forall y\in \hat \Sigma.
\label{2.3}\ee
For any $\zeta \in \G$ denote by $\zeta_Y$ the associated vector field on $Y$ and
introduce the map
\be
\cL_y: \G \to T_y Y,
\qquad
\cL_y: \zeta \mapsto \zeta_Y(y).
\label{2.4}\ee
The restricted map
\be
\bar \cL_y: \G_y^\perp \to V_y
\label{2.5}\ee
is a linear isomorphism, and we also introduce its inverse
\be
\bar A_y:= (\bar \cL_y)^{-1} : V_y \to \G_y^\perp
\label{2.6}\ee
as well as the `mechanical connection'
$A_y: T_y Y \to \G$  that extends $\bar A_y$  by zero on $H_y$.
We then define the `inertia operator' $\cI_y \in \End(\G_y^\perp)$ by requiring
\be
\eta_y( \cL_y \xi, \cL_y  \zeta  ) = \B(\cI_y \xi, \zeta)
\qquad
\forall \xi, \zeta \in \G_y^\perp.
\label{2.7}\ee
The inertia operator is symmetric and positive definite with respect to
the restriction of the scalar product $\B$ to $\G_y^\perp$.
It enjoys the equivariance property
\be
\cI_{\phi_g(y)} \circ \M_g = \M_g \circ \cI_y,
\label{2.8}\ee
where $\M_g: \G_y^\perp \to \G_{\phi_g(y)}^\perp$ is given by
$\M_g(\xi) = \Ad_g(\xi)$.

The lifted action of $G$ to the
cotangent bundle $T^*Y$ is Hamiltonian with respect to the canonical symplectic structure,
and the associated momentum map
\be
\psi: T^*Y \to \G^*
\label{2.9}\ee
is given by the formula
\be
\psi(\alpha_y) = \cL_y^*(\alpha_y)
\qquad
\forall
\alpha_y \in T^*_y Y,
\label{2.10}\ee
which implies that $\psi(\alpha_y) \in \G_y^0 \simeq \G_y^\perp$,
where $\G_y^0 \subset \G^*$ is the annihilator of $\G_y$.

\section{Classical Hamiltonian reduction}
\setcounter{equation}{0}

The geodesic motion on a Riemannian manifold $(Y,\eta)$ can be modeled
as a Hamiltonian system on $T^*Y$ and
it is an important problem to describe its symmetry reductions built on  isometric,
proper group actions.
This problem is at the moment still open,  due to difficulties originating from
the singularities of $Y/G$.
Restricting to the principal orbit type one always (independently of having or not having sections)
obtains a smooth `reduced configuration space', $\check Y_{\red}:= \check Y/G$, and
therefore the symmetry reductions of $T^* \check Y$ are much easier to characterize.
In fact, Hochgerner  \cite{Ho,Ho+} gave a general analysis of cotangent bundle
reduction under assuming
a singe isotropy type for the action on the configuration space.
The result in Theorem 3.1  below follows as the special case of his result when the
$G$-action admits sections.
For convenience, because the general case is rather involved,
we  present a direct  proof in our restricted setting.

It proves convenient to utilize the  `shifting trick' of symplectic reduction \cite{OR} in order
to characterize the (singular) Marsden-Weinstein reductions of the geodesic system
on $T^*\check Y$.
That is to say, we take a coadjoint orbit $\cO$ of $G$, such that $-\cO \subset \psi(T^*\check Y)$,
and start from the extended Hamiltonian system
\be
(\check P^{\ext}, \Omega^{\ext}, \H^{\ext})
\label{3.1}\ee
defined as follows.
The extended phase space is
\be
\check P^{\ext}:=
T^* \check Y  \times \cO =
\{ (\alpha_y,\xi)\,\vert\, \alpha_y\in T^*_y \check Y,\,y\in \check Y, \,\xi\in \cO\}
\label{3.2}\ee
endowed with the product symplectic structure,
\be
\Omega^{\ext}(\alpha_y, \xi)=
(d \theta_{\check Y})(\alpha_y)
+ \omega (\xi),
\label{3.3}\ee
where $\theta_{\check Y}$ is the canonical one-form of $T^* \check Y$ and
$\omega$ is the symplectic form of the orbit $\cO$.
The extended Hamiltonian is just the kinetic energy,
\be
\H^{\ext} (\alpha_y, \xi) := \frac{1}{2} \eta_y^*(\alpha_y, \alpha_y),
\label{3.4}\ee
where $\eta_y^*$ is the metric on $T^*_y\check Y$ corresponding to the metric
$\eta_y$ on $T_y \check Y$.
Let us denote the natural diagonal action of $G$ on $P^{\ext}$ by
$\phi^{\ext}_g$ for all $g\in G$.
This action operates by combining the cotangent lift of the original action $\phi_g$
with the coadjoint action on $\cO$, and therefore
(with $\psi$ in (\ref{2.10})) the associated
momentum map is furnished by
\be
\Psi: \check P^{\ext} \to \G^*,
\qquad
\Psi(\alpha_y, \xi)= \psi(\alpha_y) + \xi.
\label{3.5}\ee
Our problem is to describe the reduced Hamiltonian system at the value $\Psi=0$, denoted as
\be
(\check P_{\red}, \Omega_{\red}, \H_{\red})
\quad\hbox{with}\quad
\check P_{\red} = \check P^{\ext}//_0 G:= \check P^\ext_{\Psi=0}/G.
\label{3.6}\ee

Referring to the notations of Chapter 2, notice that the principal
isotropy group $K\subset G$ acts naturally on $\cO$, and
the momentum map for this action is the projection
$\G^*\ni \xi\mapsto \xi\vert_\K\in \K^*$.
Then introduce
the (singular) reduced coadjoint orbit
\be
(\cO_{\red}, \omega_{\red})
\quad\hbox{with}\quad
\cO_{\red}= \cO//_0 K \simeq (\cO \cap \K^\perp)/K,
\label{3.7}\ee
where in the last equality we identified $\G^*$ with $\G$ by means of $\B$.
The reduced orbit is a stratified symplectic space in general, i.e., a
union of smooth symplectic manifolds
(see \cite{OR}).
The reduced configuration space $\check Y_{\red}$ is equipped with the inherited
Riemannian metric $\eta_{\red}$, and we denote by
$\eta_\red^*$ the corresponding scalar product on the fibers of  $T^* \check Y_{\red}$.
Finally, let $\theta_{\check Y_{\red}}$ be the natural one-form on $T^* \check Y_{\red}$.

\bigskip
\noindent
{\bf Theorem 3.1.~\cite{Ho}}
\emph{
Suppose that the $G$-action on $(Y,\eta)$ admits sections and fix a connected component
$\check \Sigma$ of the regular elements of a section $\Sigma$.
Then, using the notations introduced above, for  any orbit $-\cO \subset \psi(T^* \check Y)$,
the reduced phase space (\ref{3.6}) can be identified as
\be
\check P_{\red} = T^* \check Y_{\red} \times \cO_{\red} = \{ (p_q, [\xi])\,\vert\,
p_q\in T_q^* \check Y_{\red},\,q\in \check Y_{\red},\, [\xi] \in \cO_{\red}\}
\label{3.8}\ee
equipped with the product (stratified) symplectic structure,
\be
\Omega_{\red}(p_q, [\xi]) = (d \theta_{\check Y_{\red}})(p_q) + \omega_{\red}([\xi]).
\label{3.9}\ee
The reduced Hamiltonian arising from the kinetic energy of
the geodesic motion on $\check Y$ reads
\be
\H_{\red}(p_q, [\xi]) = \frac{1}{2} \eta_{\red}^*(p_q, p_q) +
\frac{1}{2} \B(\cI^{-1}_{y(q)} \xi,\xi),
\label{3.10}\ee
where $\eta_{\red}$ is the induced metric on $\check Y_{\red}$,
$y(q) \in \check \Sigma$ projects to $q\in \check Y_{\red}$, $[\xi]
 = K.\xi \subset \cO\cap \K^\perp$,
and $\cI_{y(q)}\in GL(\K^\perp)$ is the $K$-equivariant inertia operator (\ref{2.7}).}

\bigskip
\noindent
\textbf{Proof.}
To describe the reduced system (\ref{3.6}), the key point is to introduce the
following submanifold $S \subset \check P^{\ext}_{\Psi=0}$,
\be
S= \{ (\alpha_y, \xi)\in \check P^{\ext}_{\Psi=0} \,\vert\, y \in \check \Sigma\,\}.
\label{3.11}\ee
All $G$ orbits in $\check P^{\ext}_{\Psi=0}$
intersect $S$  and the `residual gauge transformations' corresponding to this
submanifold
 are provided  by the group $K$.
Indeed,  an element of $G$ can map some element of
$S$ into $S$ only if it belongs to $K$,
and the elements of $K$ map each element of $S$ into $S$.
As a consequence, the reduced phase space enjoys the property
\be
\check P_{\red}= \check P^{\ext}_{\Psi=0}/G = S/K.
\label{3.12}\ee
 The reduced symplectic structure and the reduced Hamiltonian will be obtained simply
 by solving the momentum map constraint
 $\psi(\alpha_y) + \xi =0$ on $S$.
 For this, let us consider
 the orthogonal decompositions
 \be
 \alpha_y = \alpha_y^H + \alpha_y^V,
 \qquad
 \xi = \xi_\K + \xi_{\K^\perp}
 \label{3.13}\ee
 corresponding to (2.2) and to $\G=\K \oplus\K^\perp$ (2.1).
 It is easy to see that the momentum map constraint
 is completely solved on $S$ by
 \be
\xi_\K=0
\quad\hbox{and}\quad
\quad
y\in \check \Sigma,\, \alpha_y^H\in H_y^*: \hbox{arbitrary}
\label{3.14}\ee
together with the formula
\be
\alpha_y^V = - \bar A_y^*(\xi_{\K^\perp}),
\label{3.15}\ee
where we identified $\G$ with $\G^*$ by $\B$ and used the mechanical connection
$\bar A_y: V_y \to \K^\perp$ (\ref{2.6}).
Notice that $\alpha_y^H \in  H_y^*\simeq T_y^*\check \Sigma$ (\ref{2.3}) can be
 naturally regarded as belonging to the cotangent bundle
$T^* \check \Sigma$,
and the parametrization of $S$ by the variables
$(\alpha_y^H, \xi_{\K^\perp})$ yields the identification
\be
S \simeq T^* \check \Sigma \times (\cO\cap \K^\perp) = \{ (\alpha_y^H, \xi_{\K^\perp})\,\vert\,
\alpha_y^H \in T^*_y\check \Sigma,\,\,y\in \check \Sigma,\,\, \xi_{\K^\perp}\in \cO\cap \K^\perp\,\}.
\label{3.16}\ee
This is a $K$-equivariant identification since $\bar A_y^*$ is a $K$-equivariant map
and $\alpha_y^H$ is $K$-invariant since $K$ is the isotropy group of all $y\in \check \Sigma$
(and the `slice representation' is always trivial at the regular points).
Now the first important point is that in terms of the identification (\ref{3.16}) the
pull-back $\Omega^{\ext}\vert_S$ of $\Omega^{\ext}$ to $S$ becomes
\be
\Omega^{\ext}\vert_S(\alpha_y^H,\xi_{\K^\perp})= (d \theta_{\check \Sigma})(\alpha_y^H)
+ \omega\vert_{\cO\cap \K^\perp}(\xi_{\K^\perp}),
\label{3.17}\ee
where $\theta_{\check \Sigma}$ is the natural one-form on $T^* \check \Sigma$.
Since $T^* \check \Sigma$ is a model of $T^* \check Y_{\red}$,
we obtain the statement of the theorem concerning the reduced symplectic structure.
The second important point is that the restriction of the
Hamiltonian (\ref{3.4}) gives, in terms of the variables
$(\alpha_y^H, \xi_{\K^\perp})$, the following function
\be
\H^\ext\vert_{S}(\alpha_y^H, \xi_{\K^\perp})=\frac{1}{2}\eta_y^*(\alpha_y^H, \alpha_y^H)
+\frac{1}{2} \eta_y^* (\bar A_y^*(\xi_{\K^\perp}), \bar A_y^*(\xi_{\K^\perp})).
\label{3.18}\ee
The first term represents the kinetic energy of a particle on
$(\check Y_{\red}, \eta_{\red})$ modeled by the submanifold $\check \Sigma$ of $\check Y$.
It follows directly from the definitions given in Chapter 2 that the second term
can be rewritten as
\be
\eta_y^* (\bar A_y^*(\xi_{\K^\perp}), \bar A_y^*(\xi_{\K^\perp}))
= \B(\cI_y^{-1} \xi_{\K^\perp}, \xi_{\K^\perp}),
\label{3.19}\ee
which yields a well-defined function on $\check P_{\red}$ since (by (\ref{2.8}))
the inertia operator $\cI_y$ is $K$-equivariant
for $y\in \check \Sigma$.
\emph{Q.E.D.}
\medskip

\medskip
\noindent
\textbf{Remark 3.2.}
Identifying $\check Y_{\red}$ with $\check \Sigma$, as in the proof, we can also
model the reduced phase space as $T^* \check \Sigma \times \cO_{\red}$.
Moreover, we can view this as the factor space of $T^* \hat \Sigma \times \cO_{\red}$ under
the natural, diagonal  action of the finite group $W(\Sigma)$.
Here, $W(\Sigma)=N(\Sigma)/K$ acts on $\cO//_0K$ since $K$ is a normal subgroup of
$N(\Sigma)$.
The Hamiltonian gives rise to a $W(\Sigma)$-invariant function on
$T^* \hat \Sigma \times \cO_{\red}$.
Therefore, all objects belonging to the reduced system can be interpreted as
$W(\Sigma)$-invariants of a system associated with the (regular) part
of the slice $\Sigma$.
This is very familiar in the examples of (spin) Calogero type models
studied e.g.~in \cite{Res,Ho,FP1,FP2}.

\medskip
\noindent
\textbf{Remark 3.3.}
The first term of the reduced Hamiltonian (\ref{3.10}) is just the kinetic energy on
 $(\check Y_{\red},\eta_{\red})$, while the second term can be viewed as potential energy
depending also on the  `spin' variables belonging to $\cO_{\red}$.
According to (3.19),
the second term can be expressed also in terms of
 the dual of the mechanical connection, and in certain cases  the function
$q \mapsto \bar A_{y(q)}^*$
yields a so-called dynamical $r$-matrix \cite{Ho,FP1}.
In some exceptional cases \cite{KKS,FP2,FP3} $\cO_{\red}$ consists of a single point,
which means that no spin degrees of freedom appear in the reduced dynamics.

\section{Quantum Hamiltonian reduction}
\setcounter{equation}{0}

Below we first recall the standard quantization of the Hamiltonian
system $(P^\ext, \Omega^{\ext}, \H^{\ext})$
defined similarly to (\ref{3.1}) with $P^{\ext} = T^* Y \times \cO$.
Then we  impose the analogue of the `first class constraints' $\Psi=0$  on the
Hilbert space of this system to obtain a reduced quantum system.
Finally, we  comment on the relation between  the outcome  of this `first
quantize then reduce' procedure
and the structure of the reduced classical system given in Theorem 3.1.
Our consideration are close to those in the paper \cite{IT}, but in the context of
polar actions we can say more about the reduced systems.

Consider a unitary representation
\be
\rho: G \to U(V)
\label{4.1}\ee
on a finite dimensional complex Hilbert space $V$ with scalar product $(\ ,\ )_V$,
and  the associated Lie algebra representation
\be
\rho': \G \to u(V),
\label{4.2}\ee
where $u(V)$ is the Lie algebra of anti-hermitian operators on $V$.
The representation $\rho$ can be viewed as the quantum mechanical analogue of the
coadjoint orbit $\cO$ that features at the classical level.
The standard quantum mechanical analogue of $P^{\ext}$ is the Hilbert space
$L^2(Y,V, d \mu_Y)$ consisting of the $V$-valued square integrable functions on $Y$
with the scalar product
\be
(\F_1, \F_2) = \int_Y (\F_1, \F_2)_V d\mu_Y,
\label{4.3}\ee
where $d\mu_Y$ is the measure associated with the Riemannian  metric $\eta$ on $Y$.
Denote by $\Delta_Y^0$ the Laplace-Beltrami operator $\Delta_Y$ of
$(Y,\eta)$ \emph{on the  domain}
$C_c^\infty(Y,V) \subset L^2(Y,V, d\mu_Y)$ containing the smooth $V$-valued functions
of compact support.
It is well-known\footnote{See e.g.~\cite{Lands}, paragraph II.3.7, and references therein.
We do not include the  Ricci scalar in the quantum Hamiltonian, but its inclusion
would not cause any extra difficulty in our arguments.}
that $\Delta_Y^0$ is \emph{essentially self-adjoint},
and $-\frac{1}{2}$ times its
closure yields the  Hamilton operator corresponding to the classical
Hamiltonian  $\H^{\ext}$.

We need the equivariant functions  $\F\in C^\infty(Y,V)^G$ that satisfy by definition
\be
\F \circ \phi_g = \rho(g) \circ \F \qquad \forall g\in G.
\label{4.4}\ee
The space of functions $C^\infty(\Sigma,V^K)^W$ can be defined similarly, where
$V^K$ is the subspace of $K$-invariant vectors in $V$, on which $W$ acts since $W=N(\Sigma)/K$.
From now on we  assume that $\dim (V^K)>0$.
It is easy to see that the restriction of functions on $Y$ to $\Sigma$
gives rise to an injective map\footnote{This map is known  to be  surjective \cite{PT} if
$\dim(V)=1$.
It would be interesting to generalize this result, and
another important question is to find all cases for which $\dim(V)>\dim(V^K)=1$
like in the examples in \cite{EFK,Obl}.}
\be
C^\infty(Y,V)^G \longrightarrow C^\infty(\Sigma,V^K)^W.
\label{4.5}\ee
Note also that any function $\F \in C^\infty(Y, V)^G$ is uniquely
determined by its restriction to a connected, open component
$\check \Sigma\subset \Sigma$.
Then introduce the linear space
\be
\Fun(\check \Sigma, V^K):= \{ f\in C^\infty(\check\Sigma, V^K)\,\vert\,
\exists \F\in C_c^\infty(Y, V)^G, \,\, f= \F\vert_{\check \Sigma}\,\}.
\label{4.6}\ee
By its isomorphism with $C^\infty_c(Y,V)^G\subset L^2(Y, V, d\mu_Y)$,
$\Fun(\check \Sigma, V^K)$ becomes a pre-Hilbert space and we denote its closure
by $\overline{\Fun}(\check \Sigma, V^K)$.
It is not difficult to verify the natural isometric isomorphism
\be
\overline{\Fun}(\check \Sigma, V^K)\simeq L^2(Y,V,d \mu_Y)^G,
\label{4.7}\ee
and it is also worth remarking that $\Fun(\check \Sigma, V^K)$ contains
$C_c^\infty(\check \Sigma, V^K)$.

Because of (\ref{4.7}),
a natural quantum mechanical analogue of the classical Hamiltonian reduction
is obtained by taking the reduced Hilbert space to be  $\overline{\Fun}(\check \Sigma, V^K)$.
The reduced Hamilton operator results from $\Delta_Y^0$ on account of the following
simple observation.
There exists a unique linear operator
\be
\Delta_{\eff}: \Fun(\check \Sigma, V^K) \to \Fun(\check \Sigma, V^K)
\label{4.8}\ee
defined by the property
\be
\Delta_{\eff} f = (\Delta_Y \F)\vert_{\check \Sigma},
\quad\hbox{for}\quad
f= \F\vert_{\check \Sigma},
\quad
\F\in C_c^\infty(Y, V)^G.
\label{4.9}\ee
In other words,
the `effective Laplace-Beltrami operator' $\Delta_{\eff}$ is
the restriction of $\Delta_Y$ to $C_c^\infty(Y,V)^G$, which is well-defined
because the metric $\eta$ is $G$-invariant. Next we present the explicit formula
of $\Delta_{\eff}$.

\subsection{The effective Laplace-Beltrami operator}

A convenient local decomposition of the Laplace-Beltrami operator into `radial'
and `orbital' (angular)
parts is always applicable if one has a local, orthogonal `cross section'
of the $G$-orbits on a Riemannian $G$-manifold \cite{Helg}.
Upon restriction to $G$-equivariant functions,  the orbital part can be calculated explicitly,
and in our case we can apply this decomposition over $\check Y$ since we are
dealing with a polar action.
Before describing the result, we need some further notations.

Thinking of $\check \Sigma$  as the (smooth part of the) reduced
configuration space, denote the elements
of $\check \Sigma$ by $q$ and consider the $G$ orbit $G.q$ through any $q\in \check \Sigma$.
Both $\check \Sigma$ and $G.q$ are regularly embedded submanifolds of $Y$ and by their embeddings
they inherit Riemannian metrics, $\eta_{\check \Sigma}$ and $\eta_{G.q}$, from
$(Y, \eta)$.
Let $\Delta_{\check \Sigma}$ and $\Delta_{G.q}$ denote the Laplace-Beltrami operators
defined on the respective Riemannian manifolds $(\check \Sigma, \eta_{\check \Sigma})$ and
$(G.q, \eta_{G.q})$.
Introduce the smooth density function $\delta: \check \Sigma \to \bR_{>0}$ by
\be
\delta (q):= \hbox{volume of the Riemannian manifold $(G.q, \eta_{G.q})$}.
\label{4.10}\ee
Of course, the volume is understood with respect to the measure belonging to $\eta_{G.q}$.
(By the same formula, $\delta$ can also be defined as a $W$-invariant function on $\Sigma$.)
Referring to (\ref{2.7}),
let us define the function $\cJ: \check \Sigma \to \End(\K^\perp)$ by
\be
\cJ := \cI\vert_{\check \Sigma},
\label{4.11}\ee
and notice that, because of the $G$-symmetry,
the inertia operator $\cJ(q)$ carries the same information as the metric $\eta_{G.q}$.
Denote by $\{ T_\alpha\}$ and $\{ T^\beta\}$  some fixed dual bases of $\K^\perp$
with respect to the scalar product $\B$,
\be
\B(T_\alpha, T^\beta) = \delta_\alpha^\beta.
\label{4.12}\ee
In fact (see also Remark 4.3 below), one has
\be
\delta (q) = C \vert\det b_{\alpha,\beta}(q) \vert^{\frac{1}{2}}
\quad\hbox{with}\quad
b_{\alpha,\beta}(q)= \B(\cJ(q) T_\alpha, T_\beta)
\label{4.13}\ee
and a $q$-independent constant $C>0$, whose value could be given but is not important for us.

\bigskip\noindent
{\bf Proposition 4.1.} \emph{On $\Fun(\check \Sigma, V^K)$ (\ref{4.6}) the
effective Laplace-Beltrami operator
(\ref{4.9}) can be expressed in the following form:
\be
\Delta_{\eff}  =\delta^{-\frac{1}{2}} \circ \Delta_{\check \Sigma} \circ \delta^{\frac{1}{2}} -
\delta^{-\frac{1}{2}} \Delta_{\check \Sigma}(\delta^{\frac{1}{2}})
+ b^{\alpha,\beta} \rho'(T_\alpha) \rho'(T_\beta).
\label{4.14}\ee
More explicitly,  this means that for any $q\in \check \Sigma$ and $f\in \Fun(\check \Sigma, V^K)$,
one has
\be
(\Delta_{\eff} f)(q)  =\delta^{-\frac{1}{2}}(q) (\Delta_{\check \Sigma}
(\delta^{\frac{1}{2}}f))(q) -
\delta^{-\frac{1}{2}}(q)  (\Delta_{\check \Sigma}\delta^{\frac{1}{2}})(q) f(q)
+ b^{\alpha,\beta}(q) \rho'(T_\alpha) \rho'(T_\beta) f(q),
\label{4.15}\ee
where the  linear operator on $V^K$ in the last term contains
the inverse $b^{\alpha,\beta}(q) = \B(\cJ^{-1}(q) T^\alpha, T^\beta)$ of the matrix
$b_{\alpha,\beta}(q)$ (\ref{4.13}).
}

\bigskip\noindent
{\bf Proof.}
Take an arbitrary function $\F \in C^\infty(Y,V)$ and consider its restrictions
\be
f:= \F \vert_{\check \Sigma}
\quad\hbox{and}\quad
\label{4.16}\F_q:= \F\vert_{G.q}
\quad \forall q\in \check \Sigma.
\ee
Then $(\Delta_Y \F)(q)$ can be found with the aid of the following well-known formula.

\medskip\noindent
{\bf Lemma 4.2.}
\emph{
Using the above notations,
\be
(\Delta_Y \F)(q) = (\Delta_{\rad} f)(q) + (\Delta_{G.q} \F_q)(q)
\label{4.17}\ee
with the radial part of $\Delta_{\check Y}$ given by
\be
\Delta_{\rad}= \delta^{-\frac{1}{2}} \circ \Delta_{\check \Sigma} \circ \delta^{\frac{1}{2}} -
\delta^{-\frac{1}{2}} \Delta_{\check \Sigma}(\delta^{\frac{1}{2}}).
\label{4.18}\ee
}

The statement of Lemma 4.2 is quite standard (see e.g.~\cite{Helg}), and one can also verify it in
a direct manner by simply writing out the Laplace-Beltrami operator in such local coordinates,
$\{y^a\} = \{ q^i\} \cup \{ z^\alpha\}$,
around $q\in \check \Sigma$ that are composed of some coordinates
$q^i$ on $\check \Sigma$ and coordinates
 $z^\alpha$
around the origin of the coset space $G/K$ according to the $G$-equivariant diffeomorphism
\be
\check \Sigma \times G/K \simeq \check Y
\label{4.19}\ee
defined by $\check \Sigma \times G/K \ni(q, gK) \mapsto \phi_g(q)\in \check Y$.
The  coordinates can be chosen by taking advantage of the local diffeomorphism
\be
\K^\perp \ni z=z^\alpha T_\alpha \mapsto e^z K \in G/K,
\label{4.20}\ee
and the metric on $Y$ is then represented by
 $g_{a,b}(q,z) := \eta_{(q,z)}(\partial_{y^a}, \partial_{y^b})$ satisfying
\be
g_{i,j}(q,z) = g_{i,j}(q,0),
\quad
g_{i,\alpha}(q,z)=0,
\quad
g_{\alpha,\beta}(q,0)=b_{\alpha,\beta}(q)
\label{4.21}\ee
with $b_{\alpha,\beta}(q)$ in (\ref{4.13}).
On account of this block-diagonal structure, the local expression
\be
\Delta_Y \longleftrightarrow  \frac{1}{\sqrt{\g}}
\partial_{y^a} \circ \sqrt{\g} g^{ab}  \circ \partial_{y^b},
\qquad
\g:=  \vert \det g_{a,b} \vert 
\label{4.22}\ee
separates into the sum of two terms, which  yield the two terms in (\ref{4.17}).
Since $f\in \Fun(\check \Sigma, V^K)$ corresponds to $\F\in C^\infty_c(Y, V)^G$,
to prove the proposition
it is now enough to calculate $\Delta_{G.q} \F_q$ for equivariant functions
$\F_q\in C^\infty(G.q, V)^G$.
The result of this latter calculation
can presumably be also found in the literature, but one can also compute
it directly by using the exponential coordinates
on $G.q \simeq G/K$ introduced above.
One finds that
\be
(\Delta_{G.q} \F_q)(q) = b^{\alpha,\beta}(q) \rho'(T_\alpha) \rho'(T_\beta) f(q)
\quad\hbox{if}\quad
\F_q \in C^\infty(G.q, V)^G,
\label{4.23}\ee
which completes the proof of the proposition. \emph{Q.E.D.}

\bigskip
\noindent
{\bf Remark 4.3.}
The coset space $G/K$ carries a $G$-invariant Haar measure, which is unique up to a
multiplicative constant. The Haar measure is associated with a $G$-invariant
differential form of top degree on $G/K$.
This differential form is uniquely determined by its value at the origin $K\in G/K$.
Upon the identification $G/K\simeq G.q$, the origin becomes $q$, and the value
of the $G$-invariant volume form associated with the metric $\eta_{G.q}$ gives at the origin
\be
\vert \det b_{\alpha,\beta}(q)\vert^{\frac{1}{2}} (dz^1 \wedge dz^2 \wedge \cdots \wedge dz^m)_q,
\qquad
m:= \dim(G/K).
\label{4.24}\ee
Formula (\ref{4.13}) follows easily from this remark.

\subsection{The reduced quantum system}

The effective Laplace-Beltrami operator  $\Delta_{\eff}$ (\ref{4.14}) can be shown to be
essentially self-adjoint on the  domain
$\Fun(\check \Sigma, V^K) \subset \overline{\Fun}(\check \Sigma, V^K)\simeq L^2(Y,V, d\mu_Y)^G$.
In order to relate the reduced Hilbert space to the Riemannian metric on the smooth part of the
reduced configuration manifold,
$(\check Y_{\red}, \eta_{\red}) \simeq (\check \Sigma, \eta_{\check \Sigma})$,
the following lemma is needed.

\bigskip
\noindent
{\bf Lemma 4.4.}
\emph{
The complement of the dense, open submanifold $\check Y \subset Y$ of principal 
orbit type
has zero measure with respect to $d\mu_Y$.}

\bigskip
\noindent
{\bf Proof.}
It is known \cite{GOV} that the non-principal orbits of a given type fill \emph{lower-dimensional}
regular submanifolds in $Y$ and at most countably many different types of orbits can occur.
Since the measure $d\mu_Y$ is smooth, and $Y$ is second countable,
this implies (see e.g.~\cite{Knapp}, page 529) that $Y\setminus \check Y$ has measure zero.
\emph{Q.E.D.}
\bigskip

To proceed further, we also need the following integration formula:
\be
\int_{Y} (\F_1, \F_2)_V d \mu_Y = \int_{\check Y} (\F_1, \F_2)_V d\mu_{\check Y}
= \int_{\check \Sigma} (f_1, f_2)_V \delta d \mu_{\check \Sigma},
\quad
\F_i\in C^\infty_c(Y, V)^G,\,\, f_i=\F_i\vert_{\check \Sigma}.
\label{4.25}\ee
The first equality is guaranteed by Lemma 4.4.
The second equality holds since, as is standard to show, the measure on
 $\check Y \simeq \check \Sigma \times G/K$ takes the product form
 \be
 d\mu_{\check Y}= (\delta d\mu_{\check \Sigma}) \times d \mu_{G/K},
 \label{4.26}\ee
where $d\mu_{G/K}$ is the probability Haar measure on $G/K$,
$d \mu_{\check \Sigma}$ is the measure on $\check \Sigma$
associated with the Riemannian metric $\eta_{\check \Sigma}$,
the density $\delta$ is defined in (\ref{4.10});
 and $(\F_1,\F_2)_V$ is $G$-invariant.
The integration formula and the  fact that
$\Fun(\check \Sigma, V^K)$ contains $C_c^\infty(\check \Sigma, V^K)$,
in association with $C_c^\infty(\check Y, V)^G \subset C_c^\infty(Y,V)^G$,
together imply that
\be
\overline{\Fun}(\check \Sigma, V^K) \simeq L^2(\check \Sigma, V^K, \delta d\mu_{\check \Sigma}).
\label{4.27}
\ee
By transforming away the factor $\delta$ from the `induced measure' $\delta d\mu_{\check\Sigma}$,
we obtain the final result.

\bigskip
\noindent
{\bf Theorem 4.5.}
\emph{
Using the above notations, the reduction of the
quantum system defined by the closure of $-\frac{1}{2}\Delta_Y$ on
$C_c^\infty(Y,V)\subset L^2(Y,V, d \mu_Y)$ leads to the
reduced Hamilton operator $-\frac{1}{2}\Delta_\red$
given by
\be
\Delta_{\red}= \delta^{\frac{1}{2}} \circ \Delta_{\eff} \circ \delta^{-\frac{1}{2}}
= \Delta_{\check \Sigma} -
\delta^{-\frac{1}{2}} (\Delta_{\check \Sigma}\delta^{\frac{1}{2}})
+ b^{\alpha,\beta} \rho'(T_\alpha) \rho'(T_\beta).
\label{4.28}\ee
This operator is essentially self-adjoint on the dense domain
$\delta^{\frac{1}{2}}\, \Fun(\check \Sigma, V^K)$ in the reduced Hilbert space
identified as  $L^2(\check \Sigma, V^K, d\mu_{\check \Sigma})$.
}

\bigskip
\noindent
{\bf Proof.}
The multiplication operator $U: f \mapsto \delta^{\frac{1}{2}} f$
is an isometry from $L^2(\check \Sigma, V^K, \delta d\mu_{\check \Sigma})$ to
$L^2(\check \Sigma, V^K,  d\mu_{\check \Sigma})$.
Plainly, Proposition 4.1 and Lemma 4.4 imply
that
$\Delta_{\red} = U \circ \Delta_{\eff} \circ U^{-1}$ is a symmetric operator on the dense
domain $U(\Fun(\check \Sigma, V^K)) = \delta^{\frac{1}{2}}\, \Fun(\check \Sigma, V^K)
\subset L^2(\check \Sigma, V^K,  d\mu_{\check \Sigma})$.
The essential self-adjointness of $\Delta_{\red}$ can be traced back to the
essential self-adjointness
of $\Delta_Y$ on $C_c^\infty(Y,V)$.  More details on this last point are
provided in \cite{FP-ROMP}.
\emph{Q.E.D.}

\medskip
Let us now compare the result of the quantum Hamiltonian reduction given by Theorem 4.5 with
the classical reduced system in Theorem 3.1.
First, the classical
kinetic energy clearly corresponds to $-\frac{1}{2}\Delta_{\check \Sigma}$.
Formally,  the second term of
the classical Hamiltonian $\H_{\red}$ (\ref{3.10}) corresponds
the third term of $-\frac{1}{2}\Delta_{\red}$ (\ref{4.28}). This term can be interpreted as
potential energy if $\dim(V^K)=1$, otherwise it is a `spin dependent potential energy'.
As represented by the second term in (\ref{4.28}),
an extra `measure factor' appears  at the quantum level in general,
which has no trace in  $\H_{\red}$.
This term gives a constant or a non-trivial contribution to the potential energy
depending on the concrete examples.

In the quantum Hamiltonian reduction we started from the
full configuration space $Y$, while classically we have restricted our attention to
$\check Y$ from the beginning.
In some sense, the outcome of the quantum Hamiltonian reduction can
nevertheless be viewed as a quantization of
the reduced classical system of Theorem 3.1
\emph{because} $Y\setminus \check Y$ has zero measure.
However, this delicate correspondence needs further investigation
(see also  \cite{Kunst}).
The structure of the full (singular) reduced phase space $P_{\red}$
coming from $T^*Y$ should be  explored, too, since it is clear
that the reduced geodesic flows may leave $\check P_{\red}\subset P_{\red}$ in certain cases.

It should be stressed that at the abstract level, on account of the natural identifications
(\ref{4.7}) and (\ref{4.27}), the
reduced Hilbert space is simply provided by the $G$-singlets $L^2(Y,V,d\mu_Y)^G$.
In some cases (for example if $Y$ is a compact Lie group)  $\Delta_Y$ possesses
pure point spectrum that
can be determined from known results (such as the Peter-Weyl theorem)  in harmonic analysis.
In such cases finding the spectrum
of the reduced Hamilton operator
becomes a problem in branching rules, since it requires
finding the above $G$-singlets among the eigensubspaces of $\Delta_Y$.

\section{Examples related to spin Sutherland type  models}
\setcounter{equation}{0}

We here recall from \cite{HPTT,Kol} a class of important hyperpolar actions on compact Lie groups.
By subjecting them to the classical and quantum Hamiltonian reduction
as described in this paper, one may obtain, and solve, a large family of
spin Sutherland type integrable models.
Details on some of these models will be reported elsewhere.

Let $Y$ be a compact, connected, semisimple Lie group carrying the Riemannian metric
induced by a multiple of the Killing form.
Take the `reduction group' $G$ to be any \emph{symmetric subgroup} of $Y\times Y$.
That is to say, $G$ is any subgroup which is pointwise fixed by an involutive automorphism
$\sigma$ of $Y\times Y$ and  contains the connected component of
the full subgroup fixed by $\sigma$.
Consider the following action of $G$ on $Y$:
\be
\phi_{(a,b)}(y) := a y b^{-1}
\qquad
\forall  y\in Y,\,\, (a,b)\in G\subset Y\times Y.
\label{5.1}\ee
This action is known to be hyperpolar (see \cite{HPTT,Kol} and the references there).
The sections are provided by certain tori,  $A\subset Y$.
In fact, $A$ is the exponential of an Abelian subalgebra $\cA$ of the correct dimension
lying in the subspace $(T_e(G.e))^\perp$ of $T_eY$.
The underlying reasons behind the appearance of Sutherland type models in association
with these actions are the
exponential parametrization of $\Sigma=A$ together with the decomposition of the
Lie algebra of $Y$ into joint eigensubspaces of $\cA$.
One can illustrate this by the particular examples to which we now turn.

First, consider $\sigma(y_1,y_2)=(y_2,y_1)$.
Then $G= \{(a,a)\vert a\in Y\} \simeq Y$ and
the action (\ref{5.1}) is  just the  adjoint action of $Y$ on itself, for which
the sections are the maximal tori of $Y$.
The associated
spin (and in exceptional cases spinless) Sutherland models were
studied in \cite{EFK,Res}.

Second, take   any non-trivial automorphism $\theta$ of $Y$ and
set $\sigma(y_1,y_2):= (\theta(y_2), \theta^{-1}(y_1))$.
Now
\be
G= \{ (\theta(a), a) \,\vert\, a\in Y\}\simeq Y,
\ee
and (\ref{5.1}) yields the action of $Y$ on itself by $\theta$-twisted conjugations,
$\phi_{(\theta(a), a)}(y) = \theta(a) y a^{-1}$.
The interesting cases are when $\theta$ corresponds to a Dynkin diagram symmetry of $Y$.
Some of the resulting generalized spin Sutherland  models have been
described in \cite{FP1,FPtwisted}.

Third, suppose that $\theta_1$ and $\theta_2$ are two  involutive automorphisms
of $Y$ and let $K_1$ and $K_2$ be corresponding symmetric
subgroups of $Y$, i.e., $(Y,K_j)$ are symmetric
pairs for $j=1,2$.
By taking $\sigma(y_1,y_2):= (\theta_1(y_1), \theta_2(y_2))$, one obtains
\be
G= K_1 \times K_2 \subset Y\times Y,
\ee
and (\ref{5.1}) becomes the so-called Hermann action on $Y$.
Besides this action, the induced action of
$K_1$ on $Y/K_2$  is also hyperpolar.
The resulting spin Sutherland type  models have not yet been explored systematically,
apart from the case \cite{FP2} of the  isotropy action of $K_1$ on the symmetric space $Y/K_1$
arising under $K_1=K_2$.
Since they include, in fact, interesting spinless cases, we
shall return to this class of models elsewhere. It could be also worthwhile to investigate
the reduced systems  induced by other polar actions given in \cite{HPTT,Kol}.

Because of the compactness of $Y$, in the above cases the corresponding Sutherland
models involve trigonometric potential functions. Hyperbolic analogues of these models
can be derived \cite{FP3} by starting from non-compact semisimple Lie groups $Y$.
This requires a slight extension of the theory of polar actions, so as
to cover suitable actions on
pseudo-Riemannian manifolds.
Rational degenerations of the trigonometric models can be obtained by
Hamiltonian reduction in those cases
for which the $G$-action has a fixed point, $p$, by using that in those
cases the representation of $G$ on $T_pY$ inherits the polar property of the original action.
It is an important open problem whether the formalism of Hamiltonian reduction under
polar actions may be extended  in such a way
to incorporate also the  elliptic  Calogero-Sutherland type models.

\bigskip
\bigskip
\noindent{\bf Acknowledgements.}
The work of L.F. was supported in part by the Hungarian
Scientific Research Fund (OTKA grant
 T049495)  and by the EU network `ENIGMA'
(contract number MRTN-CT-2004-5652).
He thanks S. Hochgerner  and L. Verhoczki  for useful
discussions. B.G.P. is grateful to J. Harnad for hospitality in Montreal.

\end{document}